# Observation of Kondo hybridization wave in UTe$_2$

Xin Yu[1,*], Shuikang Yu[1,2,*], Zheyu Wu[3], Alexander G. Eaton[3], Andrej Cabala[4], Michal Vališka[4], Jun Li[5], Rui Zhou[1,6], Yi-feng Yang[1,6,7,#], Zhenyu Wang[2,8], Peijie Sun[1,6,7], Rui Wu[1,#]

[1]*Beijing National Laboratory for Condensed Matter Physics, Institute of Physics, Chinese Academy of Sciences, Beijing 100190, China*
[2]*Department of Physics, University of Science and Technology of China, Hefei 230026, China*
[3]*Cavendish Laboratory, University of Cambridge, Cambridge, CB3 0HE, United Kingdom*
[4]*Department of Condensed Matter Physics, Faculty of Mathematics and Physics, Charles University, Ke Karlovu 5, Prague 2, 121 16, Czech Republic*
[5]*School of Sciences, Yanshan University, Qinhuangdao 066004, China*
[6]*University of Chinese Academy of Sciences, Chinese Academy of Sciences, Beijing 100190, China*
[7]*Songshan Lake Materials Laboratory, Dongguan 523808, China*
[8]*Hefei National Laboratory, University of Science and Technology of China, Hefei 230088, China*

*These authors contributed equally to this work: Xin Yu and Shuikang Yu
#E-mail: yifeng@iphy.ac.cn, wurui@iphy.ac.cn

**Condensed matter systems with strong electronic correlations often manifest a variety of intertwined ordered phases of charge, spin, orbital and other degrees of freedom. As a prototypical strongly correlated electronic system, the Kondo lattice provides fertile soil for many fascinating quantum states, including quantum criticality[1,2], unconventional superconductivity[3], hidden order[4] and topological Kondo insulator/semimetal[5,6]. The foundation of Kondo physics lies in the hybridization between localized moments and itinerant electrons. Generally, the evolution of Kondo hybridization is characterized as a broad crossover rather than a phase transition. Thus far, an ordered hybridization phase has not been observed. Here, we use scanning tunneling microscopy (STM) to identify a translational-symmetry-breaking order of Kondo hybridization wave[7,8] (KHW) for the first time on the surface of the spin-triplet heavy-fermion superconductor UTe$_2$. The unprecedented phase of KHW manifests as a periodically modulated Fano lattice, accompanied by a commensurate charge density wave (CDW) and a pronounced energy gap opening near the Fermi level ($E_F$). This KHW-imprinted CDW has an intriguing real-space texture of complementary occupation of the heavy $f$ and conduction charges, thereby forming a Kondo superlattice. The KHW is coexistent with superconductivity in UTe$_2$, which may provide valuable insight into its controversial spin-triplet pairing symmetry and the underlying mechanism. Our first experimental evidence for an ordered hybridization state potentially sheds new light on the strong correlation physics of Kondo lattice system.**

A classic Kondo lattice contains a periodic array of localized $f$-spins antiferromagnetically coupled with itinerant conduction electrons. At low temperatures, the competition between collective Kondo hybridization and magnetic ordering gives rise to a variety of exotic quantum states such as non-Fermi liquid[9], unconventional superconductivity[3] and heavy-fermion liquid[10,11] with emergent flat bands near $E_F$. A thorough understanding of the *f*-conduction (*f-c*) hybridization is therefore necessary for grasping the natures of these intricate states. Typically, the development of coherent *f-c* hybridization throughout the whole lattice does not involve a phase transition involving an ordered hybridization state. However, theoretical exceptions have been proposed to explain the mysterious hidden order state of URu$_2$Si$_2$, leading to the hybridization wave order[8] breaking space-translational symmetry and the hastatic order[12,13] breaking time-reversal symmetry. To our knowledge, the ordered state of hybridization has not been experimentally observed. CDW could be a secondary effect of KHW[8]. Therefore, investigating heavy-fermion systems with the coexistence of CDW offers a promising route to probe KHW. Nevertheless, only very few materials therein have been reported to exhibit CDW features[14-16].

Recently, the heavy-fermion metal UTe$_2$ has emerged as a research focus in correlated electronic systems due to its intriguing superconducting characteristics, including spin-triplet superconductivity[17-20], intrinsic topological superconductivity[21-25], pair density wave (PDW)[26], *etc*. However, the pairing symmetry, one of the most attractive topics of UTe$_2$, remains highly controversial. STM measurements have revealed a distinct CDW state coexistent with superconductivity on the (0-11) surface[26-30], whereas bulk-sensitive probes such as X-ray diffraction[31,32] and elastic moduli[33] measurements didn't find its signature. This CDW detected on surface is unexpectedly sensitive to magnetic fields[27,28], suggesting that it does not come from the surface lattice reconstruction and has an origin of spin interactions. More experimental evidences are required to further understand this unusual CDW, which not only is essential for exploring the mechanism behind its exotic superconductivity with critical magnetic field much higher than the Pauli limit, but also provides a rare opportunity to study KHW. In this work, by utilizing STM, we begin with reporting new behaviors of the CDW states and then introduce the concomitant KHW on the surface of UTe$_2$.

**Commensurate CDW**

The orthorhombic crystalline structure of UTe$_2$ is composed of double trigonal-prism chains of coordinated Te atoms along the *a*-axis, with each U atom located in the prism center (Fig. 1a). Given the weak coupling between Te, the natural cleavage plane for STM measurements occurs on the (0-11) facet (grey-shaded plane in Fig. 1a). On the (0-11) surface (Fig. 1b), the Te$_1$-U-Te$_2$ chains propagate along the *a*-axis. Fig. 1c,d present an STM topographic image of the (0-11) surface and its fast Fourier transform (FFT). Usually, the topmost Te$_1$ chains are easily and clearly resolved, while the chains of Te$_2$ can only be visible in high-resolution images (inset of Fig. 1c). In the FFT, Bragg peaks are represented by the wavevectors of interchain distance ($q_c$) and periodic Te$_1$ ($q_{Te}^{1,2}$). The distance between two adjacent Te along the chain is characterized by $q_c^{Te}$. In the region enclosed by Bragg peaks, there are two sets of extra maxima marked by dashed white circles and orange triangles, each consisting of a sextet: the inner $q_i^n$ (n=1-6) and the outer $q_o^m$ (m=1-6). The positions of $q_o^m$ match with those reported in previous STM works, which were regarded as incommensurate CDWs[26,27]. The inner sextet of $q_i^n$ is rarely reported until recent STM measurements using a superconducting tip[25,34].

To visualize the energy dependence of $q_i^n$, we measure differential conductance maps d$I$/d$V$(**r**, $E$)≡$g$(**r**, $E$) in the same area as Fig. 1c. The inner sextet is visible in the FFT of $g$(**r**, $E$) at each measured energy ($E_F$ and -20 meV in Fig. 1e,f and partial other energies in Extended Data Fig. 1a-i). By extracting the intensities of linecuts along the three directions (Fig. 1d) passing $q_i^n$ (n=1,2,3) as a function of energy (Extended Data Fig. 1j-l), it is evident that none of them exhibits dispersion, indicative of a charge order. Fig. 1i shows a typical d$I$/d$V$ spectrum near $E_F$ at a temperature well below $T_c$, the superconducting coherence peaks are clearly resolved and the energy gap is estimated to be 0.2 meV (inset). The superconducting in-gap states also exhibit the same modulated features at the wavevectors of $q_i^n$ (Extended Data Fig. 2). In Refs. 25,34,35, complete inner sextet of $q_i^n$ only appears inside the superconducting gap and are interpreted as the signatures of quasiparticle interference (QPI). In contrast, our data show that it persists throughout an extensive energy range spanning at least tens of meV across $E_F$. Furthermore, the inner sextet is also detected in the normal state (Extended Data Fig. 3). Therefore, we attribute these robust modulations at $q_i^n$ to CDWs which are well established prior to superconductivity.

By comparing the positions of the inner sextet and Bragg peaks, we find that the CDW states at $q_i^n$ are commensurate with the surface crystal lattice. To illustrate this commensurate relationship more clearly, a **q**-space lattice formed by the inner sextet is overlaid on the FFT of $g$(**r**, -5 meV) (Fig. 1g). Both Bragg peaks and the outer sextet of $q_o^m$ align closely with the lattice points. From this perspective, the so-called incommensurate CDW wavevectors of $q_o^m$ identified in previous works[26,27] could be regarded as the superpositions of $q_i^n$. The wavevectors of inner sextet are given by $q_i^{1,6}$ = (-1/2$q_c$, ±1/7 $q_c^{Te}$), $q_i^{2,5}$ = (0, ±2/7 $q_c^{Te}$) and $q_i^{3,4}$ = (1/2$q_c$, ±1/7 $q_c^{Te}$). This commensurability can also be manifested in real space. In a Fourier-filtered $g$(**r**, $E$) at $q_i^n$ (n=1-6) (Fig.1h and Methods), the CDW modulation marked by white circles exactly agrees with the surface crystal lattice indicated by blue dots.

**Temperature dependence of the CDW**

As a translational-symmetry-breaking order, CDW typically emerges below a phase transition temperature $T_{cdw}$, accompanied by opening an energy gap $\Delta_{cdw}$ to reduce the density of state (DOS). To determine the values of $\Delta_{cdw}$ and $T_{cdw}$, we track the thermal evolution process of the CDW.

Fig. 2a-h show the $g$(**r**, -20meV) and their FFTs at several temperatures. Generally, the amplitudes of $q_i^n$ and $q_o^m$ are suppressed as the temperature increases. Above 7 K, although the signals of $q_i^n$ are buried in the noise, the peaks of $q_o^m$ (especially $q_o^2$) remain discernible till 10 K (Extended Data Fig. 4). As the CDW gradually weakens and eventually becomes undetectable, it proves challenging to directly determine the exact vanishing temperature. One practical way is to examine the thermal decay process

of the CDW intensity (Fig. 2k and Methods). The detected CDW signals almost decay into the background level at ~ 12 K, consistent with previous report[29].

Another way to estimate $T_{cdw}$ is to detect the temperature at which $\Delta_{cdw}$ begins to open. As shown in Fig. 2i, with the temperature decreasing from 15 K to 1.4 K, the DOS within several meV near $E_F$ consistently reduces, indicating that an energy gap progressively develops. The d$I$/d$V$ spectra at 15 K and 14 K are virtually indistinguishable (bottom-right inset of Fig. 2i). To illustrate the gap more clearly, the spectra at lower temperatures are subtracted by the 15 K background (top-left inset of Fig. 2i). A pronounced energy gap feature emerges near $E_F$. Since the gap is visible from 13 K, we infer that it begins to open at ~ 14 K. The thermal decay of the gap depth at $E_F$ is plotted in Fig. 2j. Compared with Fig. 2k, the gap depth and CDW intensity have a concurrent thermal suppression and vanish at the same temperature scale. Therefore, it is reasonable to identify the observed gap as $\Delta_{cdw}$. The measured value of $\Delta_{cdw}$ at 1.4 K is ~ 6 meV and $T_{cdw}$ is estimated to be 12-14 K.

Even though bulk-sensitive probes haven't revealed a clear signature of CDW phase transition, there indeed exist many evident anomalous behaviors with a characteristic temperature $T^{*}$[36] close to $T_{cdw}$ (Fig. 2j,k). The $a$-axis magnetic susceptibility $\chi_a$ shows a rapid increase below ~ 12 K[17] while the slope of magnetization d$M_a$/d$T$ has a minimum near 14 K at the limit of zero external field[36]. A sharp peak of the $c$-axis electrical resistivity $\rho_c$ emerges near 14 K[37]. The electronic specific heat $C_e/T$ exhibits a broad peak with a maximum around 12 K[38]. Temperature-dependent curves of the linear thermal expansion coefficients show a dip centered at ~ 13 K[39]. Muon Knight shift $K_\mu$ exhibits a departure from the universal linear scaling below 12 K[40]. Although the definite origin underlying these features has yet to be resolved, the coincidence of their characteristic temperatures with $T_{cdw}$ suggests that the formation of an unusual CDW may be responsible for or at least have a contribution to them.

Since the CDW arises below the Kondo coherence temperature of ~ 50 K derived from transport measurements[37], the coherent $f$-$c$ hybridization inevitably has an impact on its formation and may make it behave unlike ordinary CDW. Coexistence of CDW and Kondo lattice has been theoretically predicted[41-44], but the signatures have only been reported in very few heavy-fermion materials, such as $Yb_5Ir_4Si_{10}$[14] and $UPt_2Si_2$[16]. A very recent STM study on a $d$-electron heavy-fermion ferromagnet, $Fe_5GeTe_2$, demonstrates that the CDW has a strong coupling with Kondo lattice[45]. It is necessary to investigate how Kondo hybridization interacts with the CDW in $UTe_2$.

**KHW as a modulated Fano lattice**
For STM measurements, the $f$-$c$ hybridization results in two interfering tunneling channels near $E_F$, causing a Fano lineshape of the d$I$/d$V$ spectrum[46-49], which is given by

$$\frac{dI}{dV}(V) \propto \frac{(q+\varepsilon)^2}{\varepsilon^2+1}; \quad \varepsilon = \frac{(V-\varepsilon_0)}{\Gamma}, \tag{1}$$

where $\varepsilon_0$ is the resonance energy, its width $\Gamma$ can represent the hybridization strength and $q$ is the tunneling ratio between $f$ and conduction orbitals. Temperature evolution of the d$I$/d$V$ spectra with Fano fittings are shown in Extended Data Fig. 5. At temperatures above $T_{cdw}$, the spectra data agree well with the Fano lineshape. Upon cooling, the deviation near $E_F$ comes from the formation of $\Delta_{cdw}$ (Extended Data Fig. 5b). The behavior of the gap opening from a background of Fano lineshape is very similar to the case of URu$_2$Si$_2$[50,51], where the gap is induced by a second-order phase transition to the so-called hidden order, whose origin is still unknown to date. In contrast, the energy gap in UTe$_2$ corresponds to a CDW, even though decisive evidence for its bulk phase transition remains elusive.

Real-space maps of Kondo hybridization can be revealed by imaging the Fano resonance. The atomic-resolved Fano lattice at the temperature above $T_{cdw}$ is presented by maps of the parameters of $\varepsilon_0$, $\Gamma$ and $q$ (Fig. 3b-d), extracted from the Fano fittings for d$I$/d$V$ spectra at each pixel (Methods). To compare the Fano lattice with crystal structure, the schematic of surface lattice identified from the topography (Fig. 3a) is overlaid on each image. Asymmetric features of the Te$_1$-U-Te$_2$ chain structure are also visible in the Fano lattice. All three parameters show maxima around U sites, as expected for Kondo screening of localized $f$-electrons. The maxima of $\varepsilon_0$ and $q$ occur between U and Te$_1$, while the maximum of $\Gamma$ occurs near Te$_2$ sites. This asymmetry reflects the nonequivalent coupling of the U $f$-orbitals with the Te$_1$ and Te$_2$ $p$-orbitals, as revealed by theoretical calculations[52].

To investigate the interplay between the CDW and Kondo hybridization, we visualize the Fano lattice below $T_{cdw}$. Surprisingly, the maps of the Fano parameters shown in Fig. 3f-h reveal obvious real-space periodic modulations beyond the crystal lattice, in contrast to Fig. 3b-d. In the FFTs (Fig. 3j-l), besides the Bragg peaks, there are also maxima at the positions of $\mathbf{q}_i^n$ and $\mathbf{q}_o^m$, indicating that the Fano lattice develops a translational-symmetry-breaking order concomitant with the CDW. Given that these Fano parameters characterize the $f$-$c$ hybridization, the observed modulated Fano lattice intrinsically represents the KHW. As the KHW and CDW concomitantly emerge from the Fano lattice, the detected energy gap near $E_F$ (Fig. 2i) actually indicates the phase transition of these two intertwined orders. In the superconducting state, the KHWs with wavevectors of $\mathbf{q}_i^n$ and $\mathbf{q}_o^m$ are also detected (Extended Data Fig. 6).

The coexistence of KHW and CDW instantly raises a question: which one is the mother order? Intuitively, the spatial charge variation could naturally modulate the Kondo screening and thereby give rise to a KHW. Conversely, CDW could also be a secondary order accompanied with KHW at the same wavevectors[8]. Previous STM studies have shown that the CDW at $\mathbf{q}_o^m$ are sensitive to magnetic field[27,28], supporting that the unusual CDW is driven by spin interactions, which inevitably includes the Kondo hybridization.

In an ordinary Kondo lattice without KHW/CDW, the real-space modulation of hybridization follows the crystal structure (Fig. 3a-d) with the $f$ electrons mainly occupies near U sites and the conduction electrons are itinerant throughout the whole lattice. When KHW sets in, the periodically modulated hybridization beyond the crystal lattice should be inherently accompanied by a modulation of the $f$-$c$ relative charge density. To check whether the CDW has this unique KHW-imprinted feature, we next visualize the energy-dependent charge density texture near $E_F$.

**KHW-imprinted *f-c* charge density texture**

An ideal CDW at a wavevector $\mathbf{q}$ can be expressed by $\rho_{\mathbf{q}}(\mathbf{r}) \propto \cos(\mathbf{q} \cdot \mathbf{r} + \varphi_{\mathbf{q}})$. The global phase $\varphi_{\mathbf{q}}(E)$ can be extracted from the FFTs of $g(\mathbf{r}, E)$ (Methods). We choose a defect-free region to measure $g(\mathbf{r}, E)$ and ignore the minor local variations. The energy dispersion of $\varphi_{\mathbf{q}}$ at each $\mathbf{q}_i^n$ (n=1,2,3) is shown in Fig. 4a-c. By inverse FFT filtering at each $\mathbf{q}_i^n$, the corresponding real-space charge density patterns can be individually visualized (Fig. 4d-f). To demonstrate the energy-dependent spatial variations of the patterns, we plot the intensity of Fourier-filtered $g(\mathbf{r}, E)$ as a function of position (Fig. 4g-i) along the three linecuts in Fig. 4d-f. For each $\mathbf{q}_i^n$, the modulated charge density at different energies exhibits a strong spatial dependence; in other words, the CDW having an energy-dependent spatial phase $\varphi_{\mathbf{r}}(E)$, which is the real-space manifestation of $\varphi_{\mathbf{q}}(E)$. For the CDW at $\mathbf{q}_i^1$ and $\mathbf{q}_i^3$, there are obvious jumps of both $\varphi_{\mathbf{r}}$ and $\varphi_{\mathbf{q}}$ near $E_F$ and 30 meV. In the energy range from $E_F$ to ~ 30 meV, $\varphi_{\mathbf{r}}$ is very different or even nearly has a $\pi$ shift from the outside energy range. For $\mathbf{q}_i^2$, although no obvious phase jump appears, rapid variations occur near -15 meV and 30 meV. In contrast, for the high energy range below -20 meV or above 40 meV, the phase variations are much smaller for each $\mathbf{q}_i^n$.

Generally, the variations (more than $2\pi$) of $\varphi_{\mathbf{r}}$ and $\varphi_{\mathbf{q}}$ mainly occur in the low energy range from -20 meV to 40 meV. Such large phase variations of CDW around $E_F$ are highly related to the emergent heavy-fermion bands induced by coherent *f-c* hybridization. Electronic structure calculations have shown that the hybridization energy range is several tens of meV near $E_F$ and the heavy *f*-bands mainly lie slightly above $E_F$[52,53]. Studies of angle-resolved photoemission and quantum oscillation also support that the heavy *f*-bands are just higher than $E_F$[54-59]. Combined with the above energy-dependent variations of $\varphi_{\mathbf{r}}$ and $\varphi_{\mathbf{q}}$, we infer that the heavy *f*-bands mainly sit in the energy range from $E_F$ to 30 meV.

To demonstrate the composite *f-c* charge density texture at the inner sextet of $\mathbf{q}_i^n$ (n=1-6), we show the Fourier-filtered $g(\mathbf{r}, E)$ at the energy close to heavy *f*-bands (20 meV) and the energies of conduction bands (50 meV and -30 meV) in Fig. 4j-l. Energy-dependent spatial phase variations of this composite charge density texture near $E_F$ can be viewed by cross-correlation between $\mathbf{q}_i^n$-Fourier-filtered $g(\mathbf{r}, E)$ at every energy and $g(\mathbf{r}, -40$ meV$)$ (Extended Data Fig. 7). The phase variations also mainly occur from -20 meV to 40 meV, in consistent with the above analysis for each $\mathbf{q}_i^n$. The charge density texture at 20 meV is composed of periodic isolated-like spots, agreeing with the relatively localized feature of charges from heavy *f*-bands. In contrast, the honeycomb-like textures at 50 meV or -30 meV are consistent with the more extended feature of charges from conduction bands. Moreover, the charge density texture at 20 meV has a nearly inverse contrast to the other two, suggesting the modulated charge densities of heavy *f* and conduction bands tend to occupy complementary positions.

For simplicity, we tentatively term this inverse real-space texture as a spatial occupation separation (SOS) of the modulated *f-c* charge density that is strongly imprinted with KHW. In a Kondo lattice without KHW, the localized *f* and itinerant conduction electrons intrinsically carry different real-space occupation weights at a given site, so to speak, forming an *f-c* SOS on the scale of crystal structure. In the KHW state of UTe$_2$, the long-range modulated *f-c* SOS beyond the scale of underlying lattice can be equivalently regarded as a Kondo superlattice.

**Discussion**

It is worth discussing whether the KHW/CDW of UTe$_2$ is only confined on the surface. Although previous bulk-sensitive measurements didn't reveal them or sharp phase transition, there are indeed various anomalies of bulk properties near $T_{khw/cdw}$ (12K – 14 K), as mentioned above. This discrepancy combined with the sensitivity to magnetic field suggests that the CDW detected by STM is close to a purely electronic one with little or resolution-limited periodic lattice distortion (PLD). Usually, the real-space CDW texture with obvious PLD has no contrast inversion for filled or empty states[60]. Here, the contrast inversion of *f-c* charge density texture (Fig. 4j-l) and energy-dependent phase variations of the modulated charges near $E_F$ (Fig. 4 and Extended Data Fig. 7) indicate that the CDW is weakly coupled with the crystal lattice and has an electronic origin. One possible way to understand the failure in detecting bulk CDW or distinct phase transition is that KHW is the mother order. As a secondary accompanied order, CDW could be too weak[8] for bulk measurements to detect, while STM is very sensitive to the surface signal even if it is so tiny. In this context, previous bulk anomalous behaviors near $T_{khw/cdw}$ may have represented this relatively weak phase transition. Given that the KHW takes place concomitantly with the unusual CDW and the modulated *f-c* SOS is hard to develop from an ordinary CDW while could be a result of KHW, it is plausible that the CDW arises from KHW rather than vice versa. Further studies are required to resolve the precise relationship between these two intertwined orders.

As two typical U-based heavy-fermion superconductors, UTe$_2$ and URu$_2$Si$_2$[50,51] exhibit similar energy gaps of an ordered phase in the normal state. The hybridization wave was proposed previously as the hidden order in URu$_2$Si$_2$[8], which is a longstanding puzzle for decades. Our direct observation of the KHW in UTe$_2$ may provide heuristic evidence for revealing the mystery of hidden order. KHW possibly arise from the special $5f^2$-$5f^3$ configuration of U orbitals with strong spin-orbit coupling, suggesting the rich correlated physics of multi-*f*-electron heavy-fermion systems which have yet to be fully explored.

**Conclusion**

We observe a KHW for the first time on the surface of the spin-triplet superconductor UTe$_2$. Concomitant commensurate KHW and CDW develop from a Fano lattice upon opening an energy gap of ~ 6 meV. $T_{cdw/khw}$ (~ 14 K) is consistent with the characteristic temperatures of broad anomalies in bulk measurements, suggesting that the KHW/CDW is probably not only confined on the surface. Intriguingly, the modulated SOS of *f-c* charge density is strongly coupled with the hybridization, supporting the scenario that KHW is an underlying origin for CDW. Coexistence of KHW, CDW and spin-triplet superconductivity in UTe$_2$ makes it a unique platform for probing novel strong correlation physics.

# Methods

## 1. Scanning tunneling microscopy measurements

High-quality single crystals of UTe$_2$ ($T_c \sim 2.0$ K) grown by molten salt flux technique[56] were used in the STM measurements, which were carried on a homemade 10 mK STM system at the A7 experimental station of Synergetic Extreme Condition User Facility (SECUF). Pt/Ir tips were treated on polycrystalline gold. Differential conductance (d$I$/d$V$) spectra and maps $g(\mathbf{r}, E)$ were acquired using a standard lock-in technique with a voltage-modulated frequency of 827.183 Hz. Single crystal samples of UTe$_2$ were cleaved in cryogenic ultrahigh vacuum at a stage of ~ 5 K inside the dilution refrigerator and then immediately transferred into the STM head operating at 1.4 K. The temperature range of 1 K-16 K was obtained by a heater integrated onto the STM head. At each temperature, the data acquisition began after the STM setup thermally equilibrated.

## 2. Normalized CDW intensity

We quantify the temperature evolution of the CDW intensity by the following process. Firstly, we perform FFTs on the differential conductance maps of $g(\mathbf{r},-20$ meV) at various temperatures. The CDW intensity can be represented by the FFT amplitudes at the CDW wavevectors normalized by the average amplitude of three Bragg peaks ($\mathbf{q}_{Te}^1$, $\mathbf{q}_{Te}^2$, and $\mathbf{q}_c$). With the temperature increasing from 1 K to 13 K, because the FFT peaks at $\mathbf{q}_o^m$ and $\mathbf{q}_i^n$ gradually become weak and dispersed and eventually are buried in the noise at higher temperatures (Fig. 2e-h and Extended Data Fig. 4), it is impossible to measure the accurate amplitude for every wavevector. As our aim is to estimate the vanishing temperature of the CDW, we choose the representative wavevector, $\mathbf{q}_o^2$, with relatively better visibility of the peak feature at higher temperatures. The FFT amplitude of $\mathbf{q}_o^2$ is averaged over an area of 3×3 pixels below 8 K and 5×5 pixels at or above 8 K because of the dispersed signals.

## 3. Fano lineshape fitting and Fano lattice

Given that the DOS of conduction bands may exhibit a particle-hole asymmetry near $E_F$, besides the Fano factor, we add a quadratic background to get more accurate fittings to the d$I$/d$V$ spectra. The fitting equation is given by

$$\frac{dI}{dV}(V) = A_k \frac{(q+\varepsilon)^2}{\varepsilon^2+1} + aV^2 + bV + c; \quad \varepsilon = \frac{(V-\varepsilon_0)}{\Gamma}, \tag{2}$$

where $A_K$ is the amplitude of Fano factor, $\varepsilon_0$ is the resonance energy, $\Gamma$ is the width of Kondo resonance, $q$ is the $f$-$c$ tunneling probability ratio and $V$ is the sample bias in the unit of mV.

At temperatures above $T_{cdw}$ (16 K / 15 K), the spectra are in good agreement with the Fano fittings (Extended Data Fig. 5). With the temperature decreasing, because the energy gap gradually opens in the low energy range of [-6 meV, 6meV], the spectra have a deviation from the fitting curves near $E_F$. To avoid the effect of the energy gap on Fano fitting, the spectra measured at lower temperatures are fitted in the bias range

[-30 mV, -6 mV] and [6 mV, 30 mV]. The maps of Fano lattice shown in Fig. 3b-d,f-h are obtained by extracting the fitting parameters of $\varepsilon_0$, $\Gamma$, and $q$ at each pixel.

**4. Energy-dependent CDW phase**

To characterize the energy-dependent phase variations of the CDW (Fig. 4a-c), we firstly measure differential conductance maps $g(\mathbf{r}, E)$ and then perform FFT. The CDW phase $\varphi_\mathbf{q}(E)$ at each $\mathbf{q}_i^n$ (n=1,2,3) can be directly extracted from the phase component of the FFT at every measured energy. The pixel we choose is the one with maximum FFT amplitude for each $\mathbf{q}_i^n$.

**5. Fourier-filtered $g(\mathbf{r}, E)$**

For real-space visualization of the CDW at $\mathbf{q}_i^n$, we perform inverse Fourier transform of a small FFT window centered at $\mathbf{q}_i^n$. The center pixel is identified as the one with maximum FFT amplitude for each $\mathbf{q}_i^n$.

## Acknowledgements

This work was supported by National Key Research and Development Projects of China (Grant No. 2024YFA1611302), the Synergetic Extreme Condition User Facility (SECUF, https://cstr.cn/31123.02.SECUF) and CAS PIFI program (No. 2024PG0003). Y.-F.Y. and P.S. acknowledges the support from National Key R&D Program of China (No. 2022YFA1402203). Y.-F.Y. acknowledges the support from National Natural Science Foundation PG China (No. 12474136). Z.Wang acknowledges the support from Scientific Research Innovation Capability Support Project for Young Faculty (No. ZYGXQNJSKYCXNLZCXM-M25) and Quantum Science and Technology-National Science and Technology Major Project (No. 2021ZD0302802). A.G.E. acknowledges the support from the Henry Royce Institute for Advanced Materials through the Equipment Access Scheme enabling access to the Advanced Materials Characterisation Suite at Cambridge (Nos. EP/P024947/1, EP/M000524/1, and EP/R00661X/1). M.V. and A.C. acknowledges the support from the Czech Science Foundation (GACR) (No. 22-22322S).

## Author contributions

R.W. conceived and supervised the project. A.C. and M.V. grew samples. Z.Wu and A.G.E. provided and characterized samples. X.Y., S.Y. and R.W. performed STM measurements. S.Y., X.Y., R.W., Z.Wang. and J.L. carried out data analysis. R.W., P.S., Y.-F.Y., Z.Wang and R.Z. interpreted the data. R.W. wrote the paper with key contributions from Y.-F.Y. and P.S. and input from all authors. R.W. designed and constructed the homemade 10 mK STM system with X.Y.

## Competing interests

The authors declare no competing interests.

## Additional information

Correspondence and requests for materials should be addressed to Rui Wu.

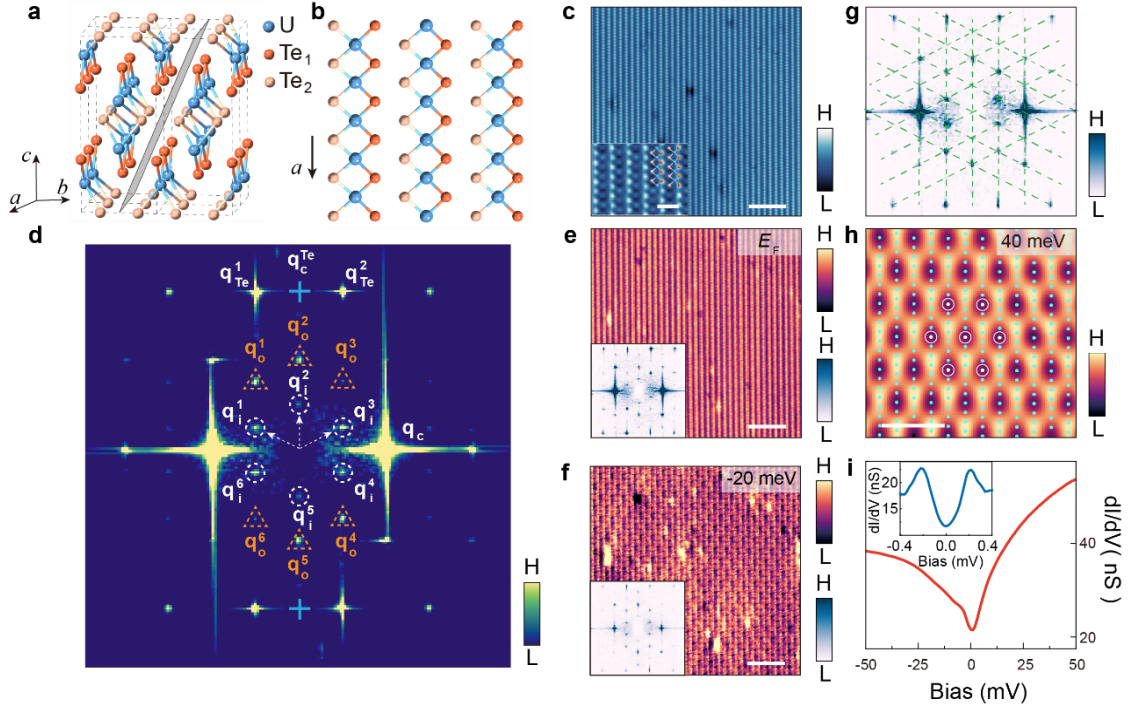

**Fig. 1 Commensurate CDW on the (0-11) surface of UTe$_2$.**
**a**, Schematic crystal structure of UTe$_2$. The grey-shaded plane shows the (0-11) facet where natural cleavage occurs. **b**, Schematic (0-11) surface crystal structure of UTe$_2$, showing the T$_1$-U-Te$_2$ chain along the *a*-axis. **c**, A typical STM topographic image on the UTe$_2$ (0-11) surface, with Te$_1$ resolved. ($I$ = 0.2 nA, $V$ = 20 mV). Inset, high-resolution topography with Te$_1$ and Te$_2$ resolved by comparing the overlaid schematic surface lattice ($I$ = 1 nA, $V$ = 10 mV). Scale bars, 5 nm (main), 1 nm (inset). **d**, FFT of the topography in **c**, with Bragg peaks ($q_{Te}^{1,2}$ and $q_c$), inner CDW sextet ($q_i^n$, dashed white circles) and outer CDW sextet ($q_o^m$, dashed orange triangles) marked. $q_c^{Te}$ (blue crosses), which is at the center between $q_{Te}^1$ and $q_{Te}^2$, marks the distance between Te along the chains. **e, f**, Differential conductance maps $g(r, E)$ of the same area as **c**, measured at $E_F$ and -20 meV ($I$ = 1 nA, $V$ = 50 mV, $T$ = 15 mK). Scale bars, 5 nm. Insets, their FFTs. **g**, FFT of $g(r, -5$ meV). The grid of green dashed lines represents the **q**-space lattice formed by $q_i^n$. Both Bragg peaks and $q_o^m$ align with the grid points, indicating the observed CDW is commensurate with the crystal lattice: $q_i^{1,6}$ = (-1/2$q_c$, ±1/7$q_c^{Te}$), $q_i^{2,5}$ = (0, ±2/7$q_c^{Te}$) and $q_i^{3,4}$ = (1/2$q_c$, ±1/7$q_c^{Te}$). **h**, Fourier-filtered $g(r, 40$ meV) at $q_i^n$ (n=1-6), showing the composite texture of the CDW with the inner sextet wavevectors ($I$ = 1.6 nA, $V$ = 60 mV). The modulated period of charge density (marked by white circles) is commensurate with the period of crystal lattice (marked by blue dots). Scale bar, 3nm. **i**, d$I$/d$V$ spectrum near $E_F$ measured at 15 mK ($I$ = 2 nA, $V$ = 50 mV), with the inset highlighting the features of superconducting gap ($I$ = 40 pA, $V$ = -1 mV, $V_{mod}$ = 0.04 meV).

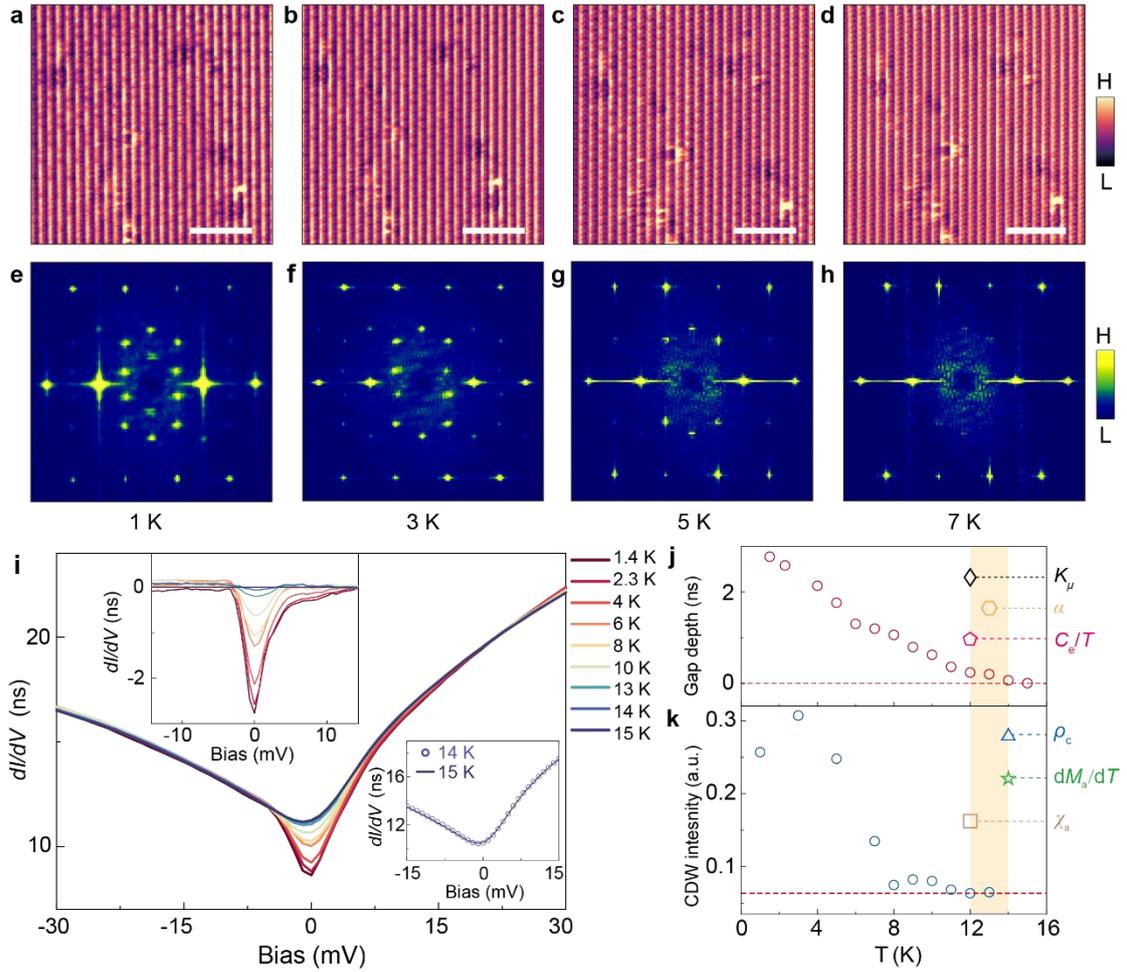

**Fig. 2 Temperature dependence of the CDW.**
**a-d**, Differential conductance maps $g(\mathbf{r}, -20\text{ meV})$ measured at 1 K **(a)**, 3 K **(b)**, 5 K **(c)** and 7 K **(d)** ($I$ = 0.7 nA, $V$ = -20 mV). Scale bars, 5 nm. **e-h**, FFTs of **a-d**, showing reduced CDW intensities at both inner and outer sextets with temperature increasing. **i**, Temperature-dependent d$I$/d$V$ spectra revealing an energy gap opening near $E_F$ ($I$ = 1 nA, $V$ = 50 mV). Each spectrum is the average of 100 spectra acquired in a 5 nm×5 nm area. Top-left inset shows the gap feature more clearly by subtracting the spectrum at 15 K. The gap is clearly visible from 13 K. Bottom-right inset compares the spectra measured at 14 K and 15 K. Their negligible difference indicates that the gap begins to open from ~ 14 K. **j**, **k** Temperature dependence of the gap depth [d$I$/d$V$ (15 K, 0 mV) - d$I$/d$V$ ($T$, 0 mV)] (**j**, red circles) and the CDW intensity (**k**, blue circles). The intensity is represented by the normalized FFT amplitudes at $\mathbf{q}_0^2$. (Methods and the FFTs above 7 K are shown in Extended Data Fig. 4). Other symbols in shaded yellow region mark the characteristic temperatures for anomalous behaviors in bulk measurements: $a$-axis magnetic susceptibility $\chi_a$ (12 K, brown square)[17], slope of $a$-axis magnetization d$M_a$/d$T$ (14 K, green pentagram)[36], $c$-axis electrical resistivity $\rho_c$ (14 K, blue triangle)[37], electronic specific heat $C_e/T$ (12 K, magenta pentagon)[38], linear thermal expansion coefficient $\alpha$ (13 K, orange hexagon)[39] and muon Knight shift $K_\mu$ (12 K, black diamond)[40].

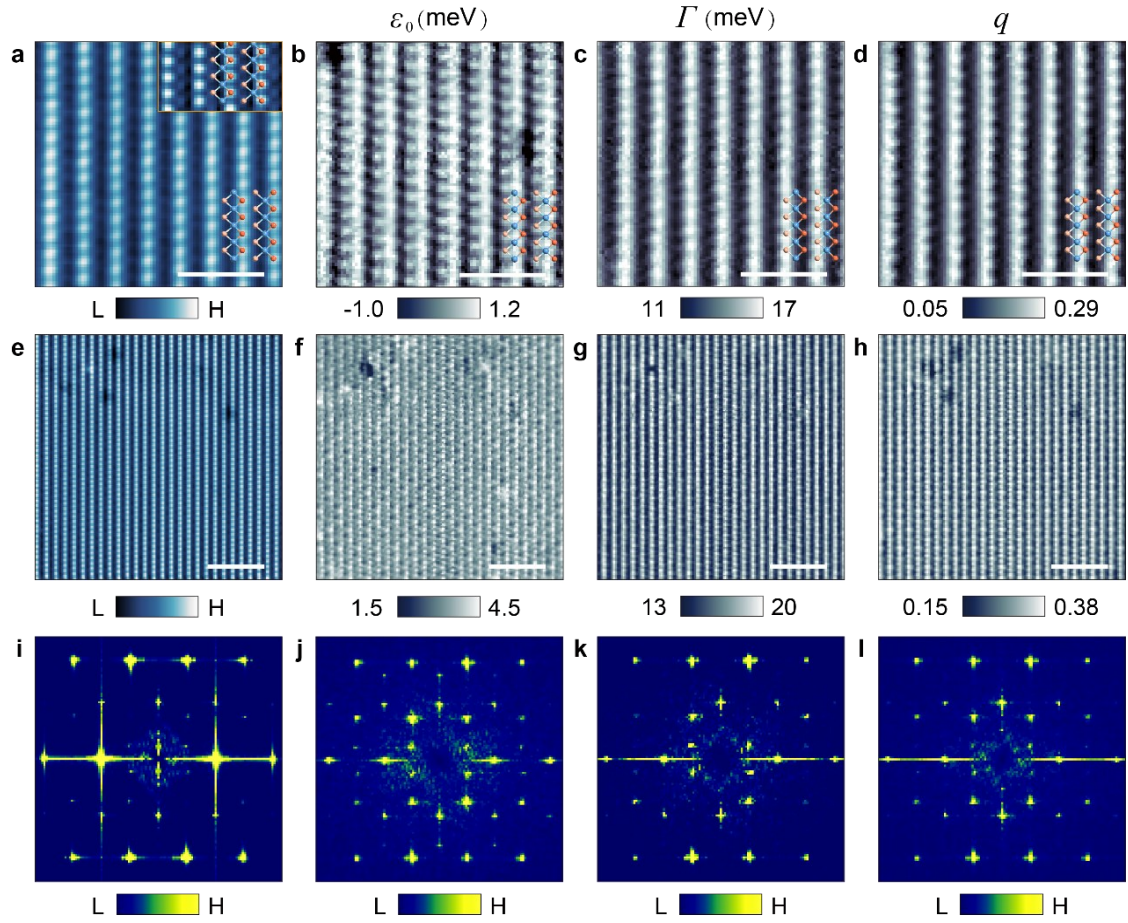

**Fig. 3 KHW as a modulated Fano lattice.**
**a-d**, Atomic-resolved Fano lattice in the same field of view measured at 16 K without CDW/KHW ($I$ = 2 nA, $V$ = 50 mV). Scale bars, 2 nm. **a**, STM topographic image of the (0-11) surface with schematic crystal lattice overlaid to show the positions for each atom. Inset, high-resolution topography. **b-d** Real-space maps of Fano parameters: resonance energy $\varepsilon_0$ **(b)**, hybridization width $\Gamma$ **(c)** and $f$-$c$ tunneling ratio $q$ **(d)**, obtained by fitting the d$I$/d$V$ spectra of Fano lineshape at each pixel (Methods). The maxima of $\varepsilon_0$ and $q$ occur between U and Te$_1$, whereas the maximum of $\Gamma$ occurs near Te$_2$. **e-h**, Atomic-resolved Fano lattice in the same field of view measured at 2.3 K ($I$ = 1 nA, $V$ = 50 mV). Scale bars, 5 nm. **e**, STM topographic image; **f-h**, Real-space maps of Fano parameters: $\varepsilon_0$ **(f)**, $\Gamma$ **(g)** and $q$ **(h)**. **i-l**, FFTs of **e-h**. Obviously, the Fano lattices of the three hybridization parameters are modulated with the same wavevectors of $\mathbf{q_i^n}$ and $\mathbf{q_o^m}$ as the CDW. Density-wave-like features on Fano lattice intrinsically represent the periodically modulated $f$-$c$ hybridization, which reveals the formation of KHW.

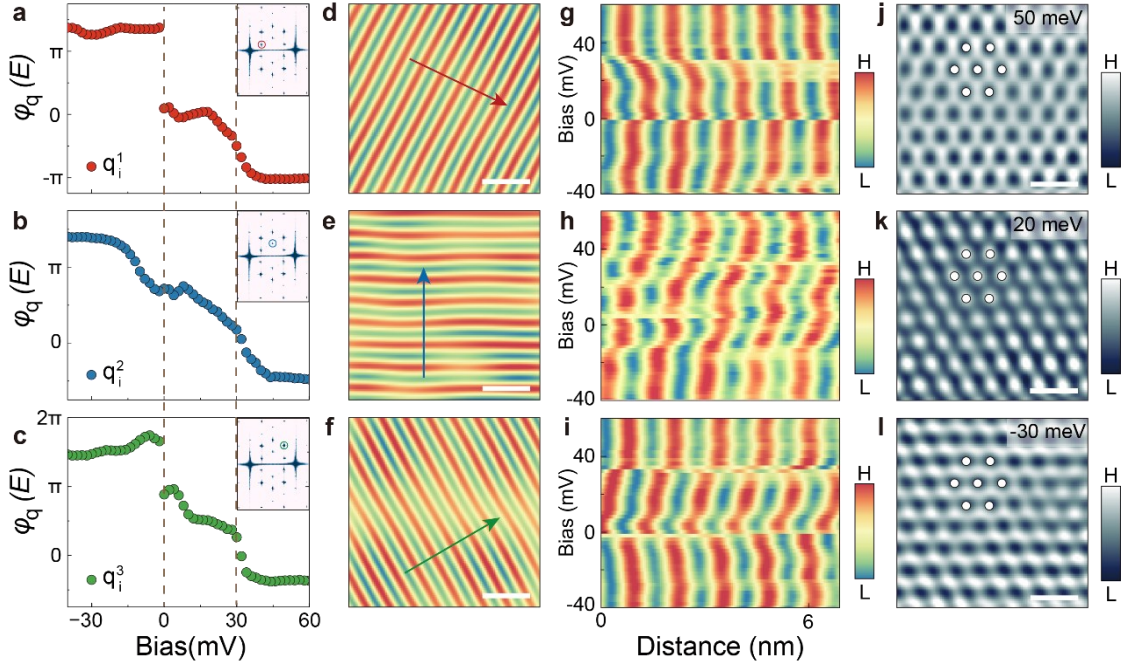

**Fig 4. Complementary *f-c* charge density texture.**

**a-c**, Energy dispersion of the CDW phase $\varphi_q(E)$ at each $\mathbf{q}_i^n$ (n=1,2,3), considering the ideal case that a CDW with wavevector **q** has a plain wave form of $\rho_\mathbf{q}(\mathbf{r}) \propto \cos(\mathbf{q}\cdot\mathbf{r}+\varphi_\mathbf{q})$. The phase at every measured energy is extracted from the FFT of a differential conductance map $g(\mathbf{r}, E)$ of a defect-free region (Methods). Inset, FFT of $g(\mathbf{r}, -6\text{ mV})$ with each $\mathbf{q}_i^n$ circled. **d-f**, Fourier-filtered $g(\mathbf{r}, -6\text{ mV})$ at each $\mathbf{q}_i^n$, showing the real-space individual CDW patterns. Minor local variations from ideal plain wave are ignored because they have no qualitatively influence for the analysis of global CDW phase $\varphi_\mathbf{q}(E)$. Scale bars, 3 nm. **g-i**, Energy-dependent intensity of Fourier-filtered $g(\mathbf{r}, E)$ at each $\mathbf{q}_i^n$ along the three linecuts in **d-f**. The real-space locations of charge density vary with energy, which can be described by a spatial phase $\varphi_\mathbf{r}(E)$. For both of $\varphi_\mathbf{r}$ and $\varphi_\mathbf{q}$, variations larger than $2\pi$ mainly occur in the low energy range from -20 meV to 40 meV. Abrupt jumps occur at $E_F$ and ~ 30 meV (dashed brown lines) for $\mathbf{q}_i^1$ and $\mathbf{q}_i^3$, and rapid variations occur near -15 meV and 30 meV for $\mathbf{q}_i^2$. The drastic CDW phase variations near $E_F$ reveal the effect of *f-c* hybridization, which makes the charge density of heavy *f* and conduction bands to occupy different real-space positions. **j-l**, Fourier-filtered $g(\mathbf{r}, E)$ at the inner sextet $\mathbf{q}_i^n$(n=1-6), showing the charge density texture at 50 meV **(j)**, 20 meV **(k)** and -30 meV **(l)**. The energy level of 20 meV is close to heavy *f*-bands and 50 meV / -30 meV is in conduction bands. White spots mark the same locations in **j-l**. Compared with the conduction bands, the modulated charge density of heavy *f*-bands tends to occupy complementary real-space positions. Scale bars, 3 nm. (*I* = 1.6 nA, *V* = 60 mV)

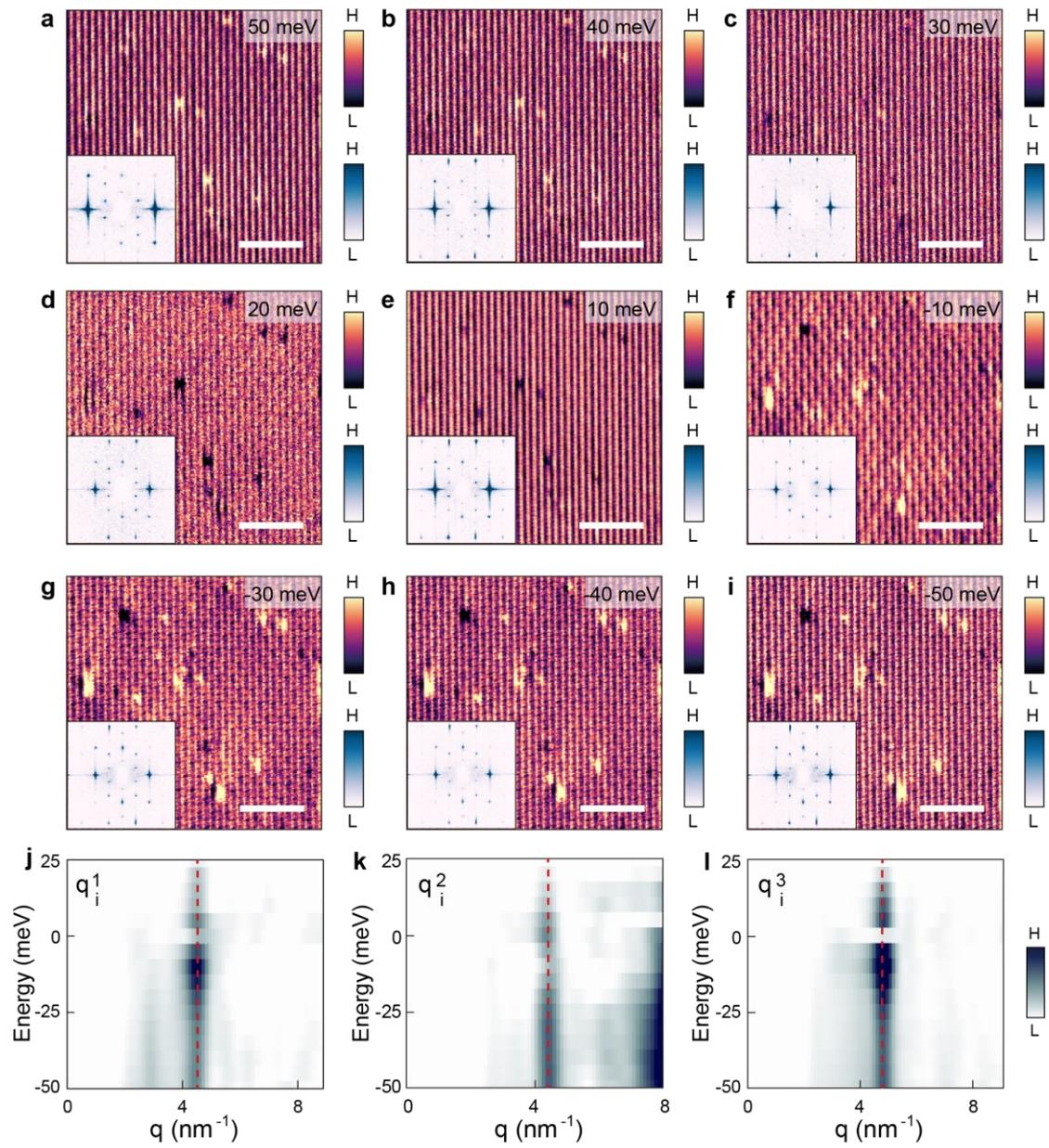

**Extended Data Fig. 1 No energy dispersion of CDW wavevectors.**
**a–i**, Differential conductance maps $g(\mathbf{r}, E)$ of the same area as Fig. 1c,e,f ($I$ = 1 nA, $V$ = 50 mV, $T$ = 15 mK). Insets, their FFTs. Scale bars, 7 nm. **j–l**, Energy-dependent amplitudes of FFT linecuts along the three directions passing $\mathbf{q}_i^n$ in Fig. 1d. Dashed red lines are guided for no energy dispersion of $\mathbf{q}_i^n$.

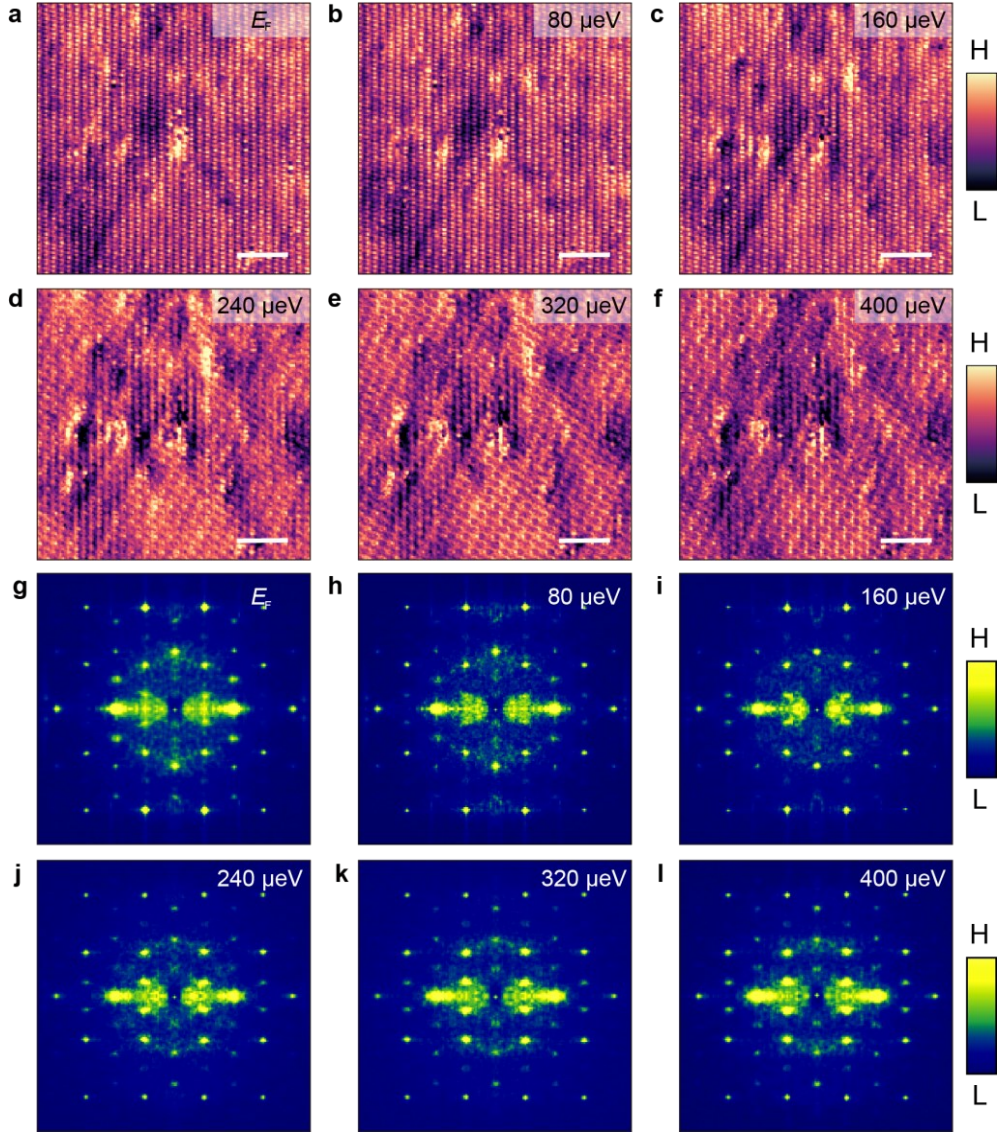

**Extended Data Fig. 2 CDW at low energies across the superconducting gap.**
**a–f**, Differential conductance maps $g(\mathbf{r}, E)$ measured in the superconducting state ($I =$ 50 pA, $V = $ -1 mV, $V_{\text{mod}} = 0.04$ meV, $T = 15$ mK). Scale bars, 5 nm. **g-l**, FFTs of **a-f**, mirror-symmetrized about the central horizontal level. Both of in-gap and out-of-gap states exhibit the same characteristic wavevectors of $\mathbf{q}_i^n$ and $\mathbf{q}_o^m$ as CDW.

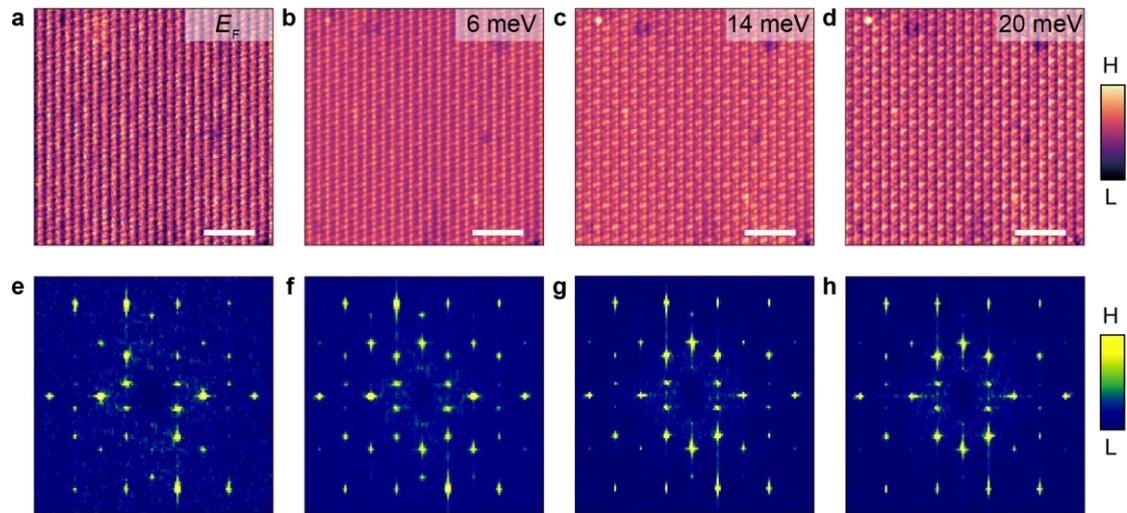

**Extended Data Fig. 3 CDW in the normal state.**
**a-d**, Differential conductance maps $g(\mathbf{r}, E)$ measured in the normal state ($I$ = 0.6 nA, $V$ = -20 mV, $T$ = 2.3 K). Scale bars, 4 nm. **e-h**, FFTs of **a-d**. The normal state exhibits the CDW wavevectors of $\mathbf{q}_i^n$ and $\mathbf{q}_o^m$.

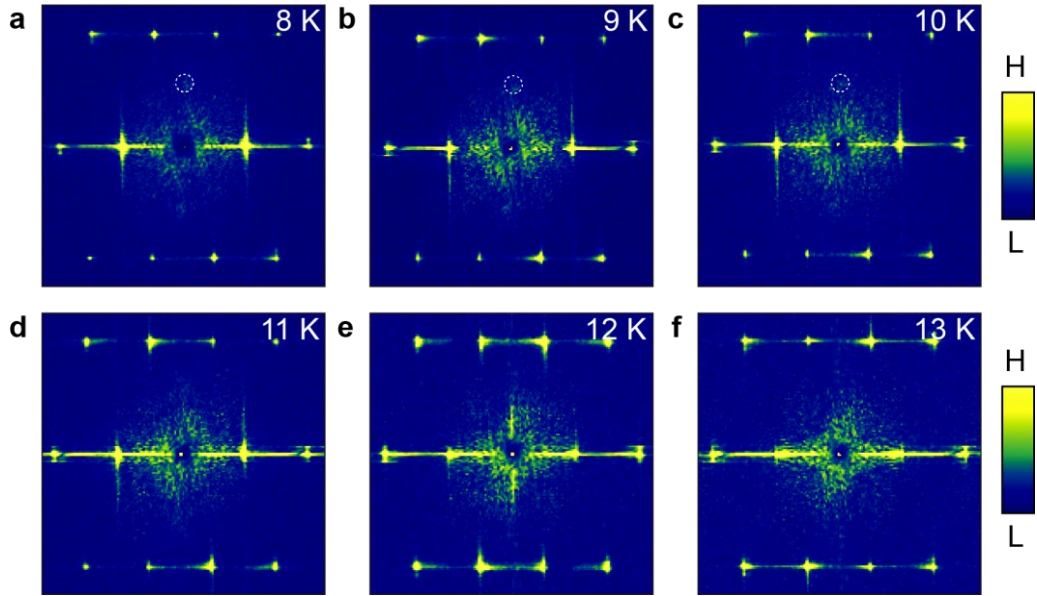

**Extended Data Fig. 4 CDW intensity decreasing with temperature**.

**a-f**, FFTs of differential conductance maps $g(\mathbf{r}, -20$ meV$)$ at different temperatures. Compared with the FFTs at lower temperatures (Fig. 2e-h), the peaks of $\mathbf{q}_o^m$ and $\mathbf{q}_i^n$ become weak and dispersed. The signals of $\mathbf{q}_o^2$ (white dashed circle) are more visible than others, so we use its amplitude to analyze the temperature dependence of the CDW intensity (Methods and Fig. 2k).

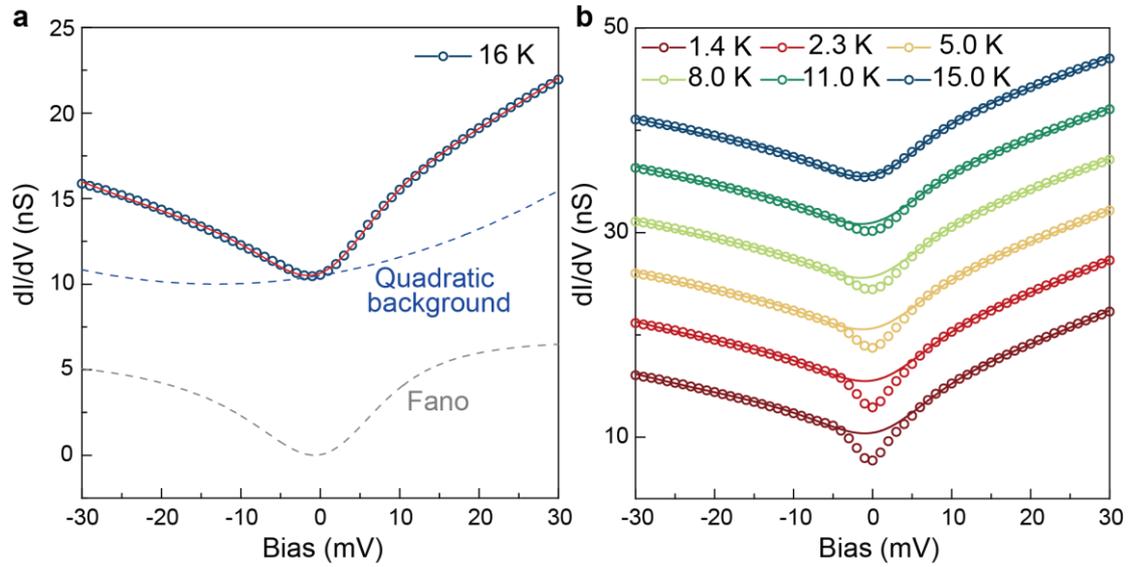

**Extended Data Fig. 5 Fano fittings to the d$I$/d$V$ spectra.**
**a**, d$I$/d$V$ spectrum measured at 16 K (dark blue circles) fitted by the red solid curve (Methods), which is composed of the quadratic background (blue dashed line) and the Fano component (grey dashed line). **b**, Temperature-dependent d$I$/d$V$ spectra (open circles) with Fano fittings (solid curves). The energy gap near $E_F$ gradually opens from the Fano resonance. Each spectrum is the average of 100 spectra acquired in a 5 nm × 5 nm area ($I$ = 1 nA, $V$ = 50 mV).

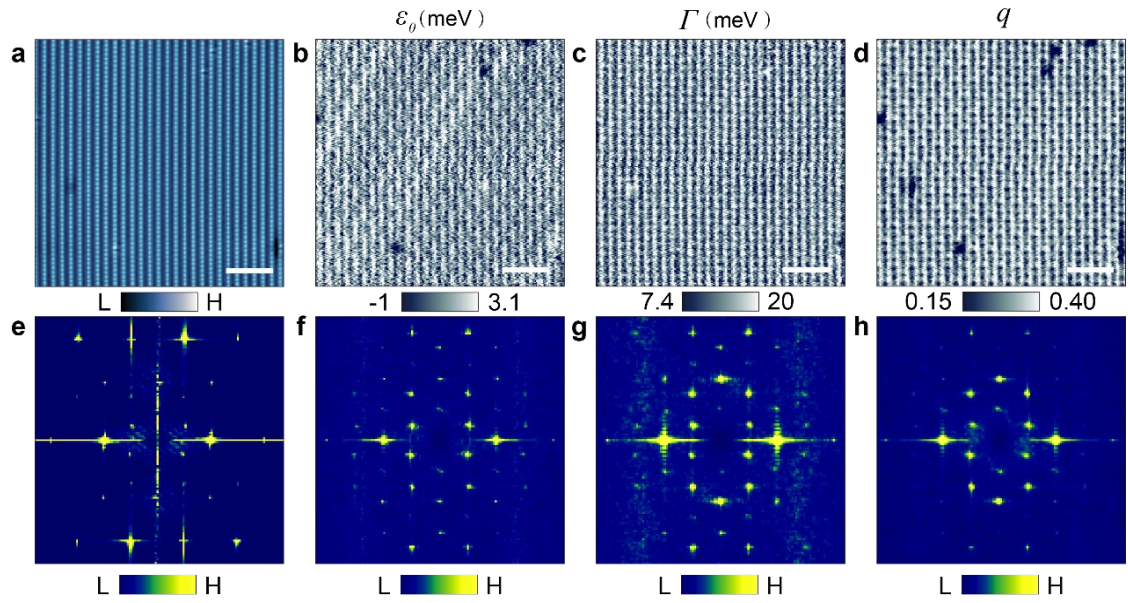

**Extended Data Fig. 6 KHW in the superconducting state.**
**a-d**, Fano lattice in the same field of view measured at 15 mK ($I$ = 1.2 nA, $V$ = 60 mV). Scale bars, 4 nm. **a**, STM topographic image. **b-d**, Real-space maps of the Fano parameters: $\varepsilon_0$ (**b**), $\Gamma$ (**c**) and $q$ (**d**). **e-h**, FFTs of **a-d**. The Fano lattice in the superconducting state has the same modulated wavevectors of $\mathbf{q}_i^n$ and $\mathbf{q}_o^m$ as the normal state (Fig. 3j-l), indicating the coexistence of KHW and superconductivity.

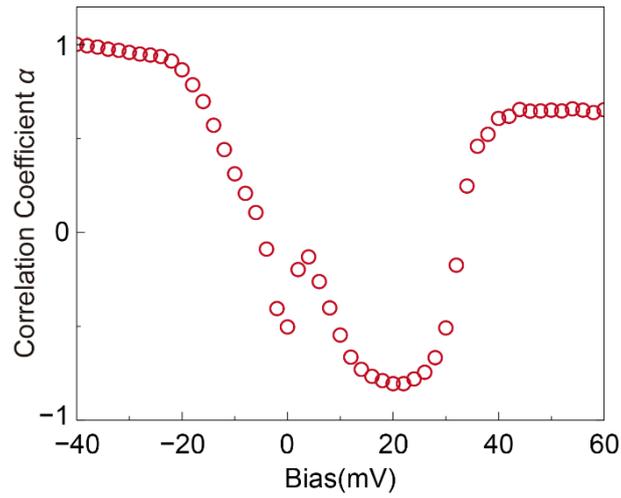

**Extended Data Fig. 7 Cross-correlation of the charge density texture near $E_F$.**
The coefficient $\alpha$ represents the cross-correlation between $\mathbf{q}_i^n$-Fourier-filtered $g(\mathbf{r}, E)$ and $g(\mathbf{r}, -40\ \text{meV})$, n=1-6. $\alpha = 1$ means the two maps are identical to each other; $\alpha = -1$ means the two maps have complete contrast inversion.